\documentclass[elsarticle.cls,superscriptaddress, notitlepage]{revtex4-1}
\usepackage{amsmath}
\usepackage{latexsym}
\usepackage{tabularx, booktabs}
\usepackage{amssymb}
\usepackage{graphics,epstopdf}
\usepackage{graphicx}
\usepackage[colorlinks=true, citecolor=blue, urlcolor=blue ]{hyperref}
\usepackage{float}
\usepackage{graphicx}
\usepackage{amsfonts}
\usepackage{color}
\usepackage{caption}
\usepackage{subcaption}
\usepackage{wrapfig, framed, caption}
    \usepackage{lipsum}%
\begin{document}

\title{Bipartite qutrit local realist inequalities and the robustness of their quantum mechanical violation}

\author{Debarshi Das}
\email{debarshidas@jcbose.ac.in}
\affiliation{Centre for Astroparticle Physics and Space Science (CAPSS), Bose Institute, Block EN, Sector V, Salt Lake, Kolkata 700 091, India}

\author{Shounak Datta}
\email{shounak.datta@bose.res.in}
\affiliation{S. N. Bose National Centre for Basic Sciences, Block JD, Sector III, Salt Lake, Kolkata 700 098, India}

\author{Suchetana Goswami}
\email{suchetana.goswami@bose.res.in}
\affiliation{S. N. Bose National Centre for Basic Sciences, Block JD, Sector III, Salt Lake, Kolkata 700 098, India}

\author{A. S. Majumdar}
\email{archan@bose.res.in}
\affiliation{S. N. Bose National Centre for Basic Sciences, Block JD, Sector III, Salt Lake, Kolkata 700 098, India}

\author{Dipankar Home}
\email{dhome@jcbose.ac.in}
\affiliation{Center for Astroparticle Physics and Space Science (CAPSS), Bose Institute, Block EN, Sector V, Salt Lake, Kolkata 700 091, India}

\begin{abstract}
Distinct from the type of local realist inequality (known as the Collins-Gisin-Linden-Massar-Popescu or CGLMP inequality) usually used for bipartite qutrit systems, we formulate a new set of local realist inequalities for bipartite qutrits by generalizing Wigner's argument that was originally formulated for the bipartite qubit singlet state. This treatment assumes existence of the overall joint probability distributions in the underlying stochastic hidden variable space for the measurement outcomes pertaining to the relevant trichotomic observables, satisfying the locality condition and yielding the measurable marginal probabilities. Such generalized Wigner inequalities (GWI) do not reduce to Bell-CHSH type inequalities by clubbing any two outcomes, and are violated by quantum mechanics (QM) for both the bipartite qutrit isotropic and singlet states using trichotomic observables defined by six-port beam splitter as well as by the spin-$1$ component observables. The efficacy of GWI is then probed in these cases by comparing the QM violation of GWI with that obtained for the CGLMP inequality. This comparison is done by incorporating white noise in the singlet and isotropic qutrit states. It is found that for the six-port beam splitter observables, QM violation of GWI is more robust than that of the CGLMP inequality for singlet qutrit states, while for isotropic qutrit states, QM violation of the CGLMP inequality is more robust. On the other hand, for the spin-$1$ component observables, QM violation of GWI is more robust for both the type of states considered.\\

\end{abstract}

\maketitle

\section{ Introduction} 

The foundational tenets and concepts of quantum mechanics (QM) significantly differ from classical ideas and intuitions. A seminal contribution to quantum concepts was provided by demonstrating quantum nonlocality through Bell's inequality \cite{Bell,CHSH} used for showing an incompatibility between quantum mechanics (QM) and the notion of local realism underpinning Bell's inequality. Soon after the discovery of Bell's inequality, a different formulation of local realist inequality was provided by Wigner \cite{Wigner}. This was based upon the assumption of the existence of joint probability distributions in the underlying stochastic hidden variable (HV) space pertaining to the occurrence of different possible combinations of outcomes for the measurements of the relevant observables, and these joint probability distributions are taken to yield all the observable marginal probabilities by satisfying the locality condition. However, Wigner's original formulation was restricted in showing the QM incompatibility with local realism for the bipartite qubit singlet states.
 
Subsequently, among the few studies using Wigner's approach are its use in the case of entangled neutral kaons \cite{Domenico,Bramon}, and a study of its implication for quantum key distribution \cite{gw}. Only recently, Wigner's formalism has been generalized for $N$-partite qubit states by deriving generalized Wigner inequalities (GWI) \cite{GWI}, and in another recent work, the temporal version of GWI,  namely, Wigner's form of the Leggett-Garg inequality has been derived \cite{wlgi}. Apart from these investigations, surprisingly, Wigner's approach has remained largely unexplored.
 
Against this backdrop, the motivation underlying the present paper is to extend the significance of Wigner's approach in the context of bipartite qutrit systems by developing a framework for local realist inequalities based on the assumption of the existence of joint probability distributions. Here it needs to be mentioned that investigations related to QM violations of local realist inequalities for arbitrary dimensional systems have steadily acquired much interest over the years \citep{t1, t2, t3, t4, t5, t6, t8, t9, t10, home1, cabello, t11}. In this context, we should also recall that qutrit systems are of special interest due to their experimental relevance in the areas of atomic and laser physics, as well as because of a number of foundational and information theoretic applications of qutrit systems \cite{bq1, bq2, bq6, bq7, bq10, bq11, bq12, bq13, bq14, bq15, bq16}.

For the purpose of probing quantum nonlocality of bipartite qutrit systems, particularly noteworthy is the QM incompatibility with local realism for bipartite qutrit isotropic states as studied by Collins et. al. \citep{cglmp} using the local realist inequality derived by them (known as the Collins-Gisin-Linden-Massar-Popescu inequality or CGLMP inequality). While experimental violation of the CGLMP inequality has been demonstrated for non-maximally entangled states of bipartite qutrits \citep{bq8}, here it needs to be mentioned that isotropic and singlet qutrit states are regarded to be particularly relevant in quantum information processing \cite{qic}. Hence in this paper, these states are used for studying the QM violation of the derived forms of GWI in the context of the six-port beam splitter and spin-$1$ component observables. Note that the forms of GWI for bipartite qutrits derived in this paper do not reduce to Bell-CHSH (Bell-Clauser-Horne-Shimony-Holt) type inequalities by clubbing any two outcomes.

An important point to stress is that the efficacy of any local realist inequality for demonstrating its incompatibility with QM is restricted in practical situations that are usually far from ideal. Hence the robustness of the QM violation of any local realist inequality in the presence of white noise in a given state is a key issue. The present paper provides a comparative study of the robustness of the QM violation of both the GWI and CGLMP inequality in the presence of white noise incorporated in the qutrit states considered. Results obtained in this paper demonstrate that for six-port beam splitter observables, the QM violation of CGLMP inequality is more robust against white noise for bipartite qutrit \textit{isotropic} states than that obtained by using GWI. On the other hand, for bipartite qutrit \textit{singlet} states, the QM violation of GWI is more robust against white noise than that pertinent to the CGLMP inequality. The corresponding calculations are also done for the spin-$1$ component observables. It is found that for both these types of states, the QM violation of GWI is more robust against white noise than that of the CGLMP inequality.


Interestingly, it may happen that the maximum QM violation of a local realist inequality is not obtained for maximally entangled qutrit state. In order to probe this, the maximum QM violation of GWI and the corresponding robustness against white noise present in the state has been calculated. It has been found that the maximum QM violation of GWI occurs for non-maximally asymmetric entangled qutrit state if one uses six-port beam splitter or spin-$1$ component observables. The maximum robustness of the QM violation of GWI against white noise present in a state is also compared with that of CGLMP inequality \cite{cglmp}.

Since all pure bipartite entangled states violate Bell-type local realist inequalities \cite{gisin}, it was believed that entanglement is equivalent to such violation. After the work of Werner \cite{werner}, it turned out that all mixed entangled states do not violate Bell-type local realist inequalities. Thus the issue of QM violations of local realism by mixed states is worth to be investigated. Motivated by this fact, we have also discussed the QM violation of GWI for some specific classes of mixed bipartite qutrit states contingent upon using six-port beam splitter as well as spin-$1$ component observables.

The plan of this paper is as follows. In the next Section II we briefly outline the original derivation by Wigner applicable for bipartite qubit singlet state. Then, in Section III, we present the derivation of GWI for bipartite qutrit systems, followed by Sections IV and V where it is shown that the derived GWI is violated by isotropic and singlet qutrit states using six-port beam splitter and spin-$1$ component observables respectively. In Sections VI and VIII, contingent upon using six-port beam splitter and spin-$1$ observables respectively, we compare the robustness of QM violations of GWI with that of CGLMP inequality for the case of isotropic and singlet qutrit states. We consider the introduction of white noise to the pure states considered in order to perform the comparative study of robustness of GWI and CGLMP inequality corresponding to the above mentioned two categories of entangled qutrit states. In Sections VII and IX we have shown the maximum QM violations of GWI using six-port beam splitter and spin-$1$ component observables respectively, and the corresponding maximum robustness of the QM violations of GWI against white noise present in the states. QM violations of GWI for mixed bipartite qutrits have been discussed in Section X. Section XI contains a summary of the results obtained in this paper and we make some concluding remarks.

\section{Recapitulating Wigner's original derivation}

In the scenario considered by Wigner \citep{Wigner}, two spin-$1/2$ particles are prepared in a singlet state and are then spatially separated. The spin components of the particles, respectively, are measured along three directions, say, $a,b$ and $c$. Then, in this context, considering the individual outcomes ($\pm 1$) of nine possible pairs of measurements, Wigner's original inequality can be derived as follows.

Assuming the locality condition and an underlying stochastic HV distribution corresponding to a quantum state specified by a wave function, one can infer in the HV space, according to the reality condition, the existence of overall joint probabilities for the individual outcomes of measuring the pertinent observables, from which the observable marginal probabilities can be obtained. Thus, corresponding to an underlying stochastic HV, say $\lambda$, one can define $p_\lambda(v_1(a),v_1(b),v_1(c);v_2(a),v_2(b),v_2(c))$ as the overall joint probability of occurrence of the outcomes, where $v_1(a)$ represents an outcome ($\pm 1$) of the measurement of the observable $a$ for the first particle, and so on. For example,  $p_\lambda(-,+,-;+,+,-)$ expresses the overall joint probability of occurrence of the outcomes $v_1(a)=-1,v_1(b)=+1,v_1(c)=-1$ for the first particle, and $v_2(a)=+1,v_2(b)=+1,v_2(c)=-1$ for the second particle. Then, the joint probability, say, $v_1(a)=+1$ and $v_2(b)=+1$ for the first and the second particle respectively can be written, using the perfect anti-correlation property of the singlet state, as $ p_\lambda(a+,b+)=p_\lambda(+,-,+;-,+,-)+p_\lambda(+,-,-;-,+,+)$. Similarly, writing $p_\lambda(c+,b+)$ and $p_\lambda(a+,c+)$ as marginals, and assuming non-negativity of the overall joint probability distributions in the HV space, it can be shown that
\begin{equation} 
\label{equ0}
p_\lambda(a+,b+)\leq p_\lambda(a+,c+)+p_\lambda(c+,b+)
\end{equation} 
Subsequently, by integrating over the hidden variable space for an arbitrary distribution, one can obtain the original form of Wigner's inequality
\begin{equation} 
\label{equ1}
p(a+,b+)\leq p(a+,c+)+p(c+,b+).
\end{equation}
where $p(a+,b+)$ is the observable joint probability of getting $+1$ for both the outcomes if the observables $a$ and $b$ are measured on the first and the second particle respectively, and so on.

If the respective angles between $a$ and $b$, $a$ and $c$, $b$ and $c$ are $\theta_{12},\theta_{13}$ and $\theta_{23}$, then substituting the QM expressions for the relevant joint probabilities in the inequality given by Eq.(\ref{equ1}) one obtains $\frac{1}{2} \sin^2(\theta_{12}/2) \leq \frac{1}{2} \sin^2(\theta_{13}/2) +\frac{1}{2} \sin^2(\theta_{23}/2)$ - a relation which is not valid for arbitrary values of $\theta_{12},\theta_{13}, \theta_{23}$. This shows an incompatibility between QM and Wigner's form of inequality given by Eq.(\ref{equ1}), restricted for the singlet state in the bipartite case. Note that, the above argument is within the framework of stochastic HV theory, subject to the locality condition, and the notion of determinism has not been used here.

\section{Generalized Wigner inequalities for bipartite qutrit systems}

Now, in order to generalise the above argument for deriving GWI for arbitrary bipartite qutrit systems, we proceed as follows. Note that in the following derivation we are not using the assumption of perfect anti-correlation embodied in the singlet states that was used in Wigner's original derivation. Let us consider that pairs of trichotomic observables $a^1$ or $a^2$ and $b^1$ or $b^2$ are measured on the first and the second particle respectively. We assume an underlying HV distribution given by $\rho(\lambda)$ such that for $3^4$ possible combinations of pairs of outcomes, each such pair of outcomes occur with a certain probability in the HV space. Thus, corresponding  to an underlying stochastic HV, say $\lambda$, one can define $p_{\lambda}(v_1(a^1), v_1(a^2); v_2(b^1), v_2(b^2))$ as overall joint probability of occurrence of the outcomes, where $v_1(a^1)$ represents an outcome ($+1$, or $0$, or $-1$) of the measurement of the observable $a^1$ for the first particle, and so on. For example, $p_{\lambda}(+,0;-,+)$ expresses the overall joint probability of occurrence of the outcomes $v_1(a^1) = +1$ and $v_1(a^2) = 0$ for the first particle and $v_2(b^1) = -1$ and $v_2(b^2) = +1$ for the second particle. Then, consistent with the locality condition, the joint probability of, say $v_1(a^1) = 0$ and $v_2(b^1)=-$ for the first and second particle, respectively, can be obtained as a marginal of the overall joint probabilities in the HV space, given by the following expression
\begin{equation}
p_{\lambda}(a^1 0 ,b^1 -) = \sum_{v_1(a^2)=+,0,-} \sum_{v_2(b^2)=+,0,-} p_{\lambda}(0, v_1(a^2); -, v_2(b^2)) \nonumber
\end{equation}
Similarly, writing $p_{\lambda}(a^1 -, b^1 -)$, $p_{\lambda}(a^2 0, b^1 -)$, $p_{\lambda}(a^2 -, b^1 -)$, $p_{\lambda}(a^1 0, b^2 -)$, $p_{\lambda}(a^1 -, b^2 -)$, $p_{\lambda}(a^2 +, b^2 +)$ and $p_{\lambda}(a^2 +, b^2 0)$ as marginals, and assuming non-negativity of the overall joint probability distributions in the HV space, it can be shown that
\begin{equation}
p_{\lambda}(a^1 0, b^1 -)  - p_{\lambda}(a^2 0, b^1 -) - p_{\lambda}(a^2 -, b^1 -)- p_{\lambda}(a^1 0, b^2 -) - p_{\lambda}(a^1 -, b^2 -) - p_{\lambda}(a^2 +, b^2 +) - p_{\lambda}(a^2 +, b^2 0) + p_{\lambda}(a^1 -, b^1 -) \leq 0 \nonumber
\end{equation}
Subsequently, integrating over the HV space for an arbitrary distribution, one can obtain the following form of GWI for bipartite qutrit systems:
\begin{equation}
\label{equgwi}
p(a^1 0, b^1 -) - p(a^2 0, b^1 -) - p(a^2 -, b^1 -) - p(a^1 0, b^2 -)  - p(a^1 -, b^2 -) - p(a^2 +, b^2 +) - p(a^2 +, b^2 0)  + p(a^1 -, b^1 -) \leq 0
\end{equation}
Similarly, other forms of $8$-term GWI can be derived by using various combinations of the observable joint probabilities. Such forms of GWI (including the above form mentioned in Eq.(\ref{equgwi})) can be expressed by the following two inequalities:
\begin{equation}
p(a^1=m_1, b^1 = m_2) - p(a^2 = m_1, b^1 = m_2) - p(a^2 = m_1, b^1 = m_1) - p(a^1 = m_1, b^2 = m_2) - p(a^1 = m_1, b^2 = m_1)  \nonumber
\end{equation}
\begin{equation}
\label{equgwi2}
 - p(a^2 = m_3, b^2 = m_3) - p(a^2 = m_2, b^2 = m_3)  + p(a^1 = m_1, b^1 = m_1) \leq 0
\end{equation}
and
\begin{equation}
p(a^1 = m_1, b^1 = m_2) - p(a^2 = m_1, b^1 = m_2) - p(a^2 = m_2, b^1 = m_2) - p(a^1 = m_1, b^2 = m_2) - p(a^1 = m_2, b^2 = m_2)  \nonumber
\end{equation}
\begin{equation}
\label{equgwi3}
- p(a^2 = m_3, b^2 = m_3) - p(a^2 = m_3, b^2 = m_1)  + p(a^1 = m_2, b^1 = m_2) \leq 0
\end{equation}
There are six permutations of the set $\{ m_1, m_2, m_3 \}$, namely: $(+1, 0, -1)$, $(+1, -1, 0)$, $(0, +1, -1)$, $(0, -1, +1)$, $(-1, +1, 0)$ and $(-1, 0, +1)$, which produce twelve GWI from the inequalities (\ref{equgwi2}) and (\ref{equgwi3}) (including the GWI mentioned in Eq.(\ref{equgwi})).
Now, interchanging $a \leftrightarrow a'$, or $b \leftrightarrow b'$, or interchanging both other three sets of twelve such $8$-term GWI can be obtained for the bipartite qutrit system. QM violations of all the aforementioned GWIs for bipartite qutrit system are quantified by the positive value of the left hand side of each inequality.

Here it needs to be stressed that this set of inequalities is such that none of these inequalities can be reduced to equivalent classes of Bell-CHSH inequalities by grouping any two outcomes (for details, see  Appendix A).

\section{QM violations of GWI by bipartite qutrit isotropic and singlet states using six-port beam splitter} 

The phenomenon of spontaneous parametric down-conversion can be used to obtain an optical analog of the maximally entangled state for two correlated spins of arbitrary magnitudes \cite{spdc}. Next, to make measurements of nondichotomic observables, it is experimentally more convenient to use six-port (or, multi port) beam splitters than spin component observables.

The properties of the unbiased six-port beam splitter (three input and three output ports) have been demonstrated in detail in several works \cite{tt1, tt2, cglmp, six, six1, six2, six3}. One considers the following settings: first the two parties apply unitary operations on each subsystem with non-zero diagonal terms equal to $e^{i \phi_a(j)}$ and $e^{i \varphi_b(j)}$ for the first and second particle respectively, and all off-diagonal terms being equal to zero. These unitary operations are denoted by $U(\vec{\phi_a})$, where $\vec{\phi_a} \equiv [\phi_a(0), \phi_a(1), \phi_a(2)]$ for the first particle  and $U(\vec{\varphi_b})$, where $\vec{\varphi_b} \equiv [\varphi_b(0), \varphi_b(1), \varphi_b(2))]$ for the second particle. The freedom of choice of the measurement of both the particles is given by this unitary transformation. Then, a discrete Fourier transformation $U_{FT}$ is carried out on the first particle and $U_{FT}^*$ is carried out on the second particle. The matrix element of the discrete Fourier transformation is given by, ${(U_{FT})}_{jk}$ = $exp[(j-1)(k-1)i2\pi /3]$ and finally measurement is done in the basis in which the initial shared state is prepared. Here the observables $a^1$, $a^2$, $b^1$ and $b^2$ denote unitary transformations $U(\vec{\phi_{a^1}})$, $U(\vec{\phi_{a^2}})$, $U(\vec{\varphi_{b^1}})$ and $U(\vec{\varphi_{b^2}})$ respectively, where $\vec{\phi_{a^1}} \equiv [\phi_{a^1}(0), \phi_{a^1}(1), \phi_{a^1}(2)]$, $\vec{\phi_{a^2}} \equiv [\phi_{a^2}(0), \phi_{a^2}(1), \phi_{a^2}(2)]$, $\vec{\varphi_{b^1}} \equiv [\varphi_{b^1}(0), \varphi_{b^1}(1), \varphi_{b^1}(2))]$ and $\vec{\varphi_{b^2}} \equiv [\varphi_{b^2}(0), \varphi_{b^2}(1), \varphi_{b^2}(2))]$.

\subsection{QM violation of GWI for bipartite qutrit isotropic state using six-port beam splitter}

Let us consider the pure isotropic qutrit state given by
\begin{eqnarray}
\label{state1}
\vert\psi_1\rangle = \frac{\vert 00 \rangle + \vert 11 \rangle + \vert 22 \rangle}{\sqrt{3}}
\end{eqnarray}
where $\vert 0 \rangle$, $\vert 1 \rangle$ and $\vert 2 \rangle$ are three mutually orthonormal states. In the case of six-port beam splitter, each of these states defines the state of photon passing through one of the three input ports or one of the three output ports of the six-port beam splitter. On the other hand, in the case of spin-$1$ component observables, $\vert 0 \rangle$, $\vert 1 \rangle$ and $\vert 2 \rangle$ are the eigenstates of spin angular momentum operator along z-direction corresponding to the eigenvalues $+1$, $0$ and $-1$ respectively (assuming $\hbar = 1$).

If measurements defined by the six-port beam splitter are performed on two particles of the state given by Eq.(\ref{state1}), the left hand side of the GWI given by Eq.(\ref{equgwi}) becomes
\begin{equation}
\label{gwilhs11}
W = \frac{1}{27}[-12 - 2 ( \sum_{i=1}^{2} \sum_{j=1}^{2}  \sum_{k=0}^{2} (-1)^{\delta_{i+j,2}} \big[ sin(\frac{\pi}{6}-\alpha_{ik}-\beta_{jk}) \big]
\end{equation}
where, $\delta_{i+j,2}$ is the Kronecker delta function; $\alpha_{ik}=\big[ \phi_{a^i}(k)-\phi_{a^i}(k+1 \text{ mod } 3) \big]$; $\beta_{jk} = \big[ \varphi_{b^j}(k)-\varphi_{b^j}(k+1 \text{ mod } 3) \big]$. In order to obtain the maximum QM violation of GWI given by Eq.(\ref{equgwi}) for pure isotropic qutrit state, we have to maximize the right hand side of Eq.(\ref{gwilhs11}), where $0 \leq \phi_{a^i} (j) \leq 2 \pi$ and $0 \leq \varphi_{b^i} (j) \leq 2 \pi$ ($i=1,2$; and $j=0,1,2$). For this maximization we have used a numerical procedure (analytical maximization is too difficult, because one has to find the global maximum of a twelve-variable function defined on some bounded twelve-dimensional domain) based on the downhill simplex method (so-called Nelder-Mead method or amoeba method) \cite{nm}. If the dimension of the domain of a function is $D$ (in our case $D=12$), the procedure first randomly generates $D+1$ points. In this way it creates the vertices of a starting the simplex. Next it calculates the value of the function at the vertices and starts exploring the space by stretching and contracting the simplex. In every step, when it finds vertices where the value of the function is higher than in others, it goes in this direction  \cite{nm}.
Following this numerical procedure we observe that for the set of measurement settings ($\phi_{a^1}(0)$, $\phi_{a^1}(1)$, $\phi_{a^1}(2)$, $\phi_{a^2}(0)$, $\phi_{a^2}(1)$, $\phi_{a^2}(2)$, $\varphi_{b^1}(0)$, $\varphi_{b^1}(1)$, $\varphi_{b^1}(2)$, $\varphi_{b^2}(0)$, $\varphi_{b^2}(1)$, $\varphi_{b^2}(1) $)= (4.62, 3.02, 3.93, 2.46, 1.80, 0.81, 0.43, 4.80, 4.64, 4.01, 3.04, 0.98) in radians, the maximum QM violation of the GWI (\ref{equgwi}) occurs, and the magnitude of this maximum violation is found to be 0.12949.

\subsection{QM violation of GWI for bipartite qutrit singlet state using six port beam splitter}

Similar to the way discussed above, it can be shown that if the trichotomic measurements labeled by ($a^1; a^2; b^1; b^2$) denoting observables using six-port beam splitter are performed on the $3\otimes 3$-dimensional pure singlet state given by
\begin{eqnarray}
\label{state2}
\vert\psi_2\rangle = \frac{\vert 02 \rangle - \vert 11 \rangle + \vert 20 \rangle}{\sqrt{3}}
\end{eqnarray}
Following the numerical procedure based on the downhill simplex method \cite{nm} as described earlier, we obtain that the left hand side of the GWI given by, Eq.(\ref{equgwi}) has the maximum value 0.12949 for the measurement settings ($\phi_{a^1}(0)$, $\phi_{a^1}(1)$, $\phi_{a^1}(2)$, $\phi_{a^2}(0)$, $\phi_{a^2}(1)$, $\phi_{a^2}(2)$, $\varphi_{b^1}(0)$, $\varphi_{b^1}(1)$, $\varphi_{b^1}(2)$, $\varphi_{b^2}(0)$, $\varphi_{b^2}(1)$, $\varphi_{b^2}(1) $)= (4.05, 0.11, 4.45, 3.02, 0.03, 2.47, 3.53, 1.87, 2.50, 6.20, 0.17, 6.13) (in radian) corresponding to the maximum QM violation of the GWI. Numerical calculations show that GWI mentioned in Eq.(\ref{equgwi}) gives the maximum QM violation for both bipartite qutrit isotropic state and bipartite qutrit singlet state among all the GWIs derived in this paper. Henceforth, we would, therefore, consider only the GWI given by Eq.(\ref{equgwi}) in case of observables using six-port beam splitters.

\section{QM violations of GWI by bipartite qutrit isotropic and singlet states using spin-$1$ component observables} 

Let us assume, $a^i$ denotes measurements of spin component of the first particle in the directions $\hat{n_i^a} = sin\theta_i^{a} cos\phi_i^{a} \hat{x} + sin\theta_i^{a} sin\phi_i^{a} \hat{y} + cos\theta_i^{a} \hat{z}$ ($i=1,2$). Similarly, $b^j$ denotes measurements of spin component of the second particle in the directions $\hat{n_j^{b}} = sin\theta_j^{b} cos\phi_j^{b} \hat{x} + sin\theta_j^{b} sin\phi_j^{b} \hat{y} + cos\theta_j^{b} \hat{z}$ ($j=1,2$), where $\theta_i^a$, $\theta_j^b$ ($i,j=1,2$) are the polar angle; $\phi_i^a$, $\phi_j^b$  ($i,j=1,2$) are the azimuthal angle; $\hat{x}$, $\hat{y}$, and $\hat{z}$ are the unit vectors in Cartesian coordinates.

\subsection{QM violation of GWI for bipartite qutrit isotropic state using spin-$1$ component observables}

If the measurements of spin-$1$ components in arbitrary directions are performed on the isotropic state (\ref{state1}), the left hand side of the GWI given by Eq.(\ref{equgwi}) becomes
\begin{equation}
\label{gwilhs1}
\begin{split}
& W = \sum_{i=1}^{2} \sum_{j=1}^{2} \frac{(-1)}{12} (-1)^{\delta_{i+j,2}} \Big[ sin^{2}( \theta_i^a + \theta_j^b ) + 2 (1 - sin \theta_i^a sin \theta_j^b ) cos \theta_i^a cos \theta_j^b + 2 (1 - cos \theta_i^a cos \theta_j^b ) cos (\phi_i^a + \phi_j^b ) sin \theta_i^a sin \theta_j^b \\&
 + sin^2 (\phi_i^a + \phi_j^b ) sin^2 \theta_i^a  sin^2 \theta_j^b + 2 \Big]
\end{split}
\end{equation}
Here, $\delta_{i+j,2}$ is the Kronecker delta function; $0 \leq \theta_i^a \leq  \pi$, $0 \leq \theta_j^b \leq  \pi$, $0 \leq \phi_i^a \leq  2 \pi$ and $0 \leq \phi_j^b \leq  2 \pi$ ($i,j = 1,2$). Following the numerical procedure based on the downhill simplex method \cite{nm} as described earlier, it has been observed that for the set of measurement settings ($\theta_1^a,\phi_1^a;\theta_2^a,\phi_2^a;\theta_1^b,\phi_1^b;\theta_2^b,\phi_2^b$)= (1.52, 3.88; 2.60, 3.84; 0.03, 0.76; 1.08, 5.56) in radians, the maximum QM violation of the GWI (\ref{equgwi}) occurs, and the magnitude of this maximum violation is found to be 0.12077.

\subsection{QM violation of GWI for bipartite qutrit singlet state using spin-$1$ component observables}

Similar to the way discussed above, using the numerical procedure based on the downhill simplex method \cite{nm}, it can be shown that if the trichotomic measurements using spin-$1$ component observables are performed on the $3\otimes 3$-dimensional pure singlet state given by Eq.(\ref{state2}), the left hand side of the GWI given by Eq.(\ref{equgwi}) has the maximum value 0.12077 for the measurement settings ($\theta_1^a,\phi_1^a;\theta_2^a,\phi_2^a;\theta_1^b,\phi_1^b;\theta_2^b,\phi_2^b$) = (1.09, 0.05; 0.02, 0.01; 0.52, 3.19; 0.56, 0.05) (in radian) corresponding to the maximum QM violation of the GWI. Numerical calculations based on the downhill simplex method \cite{nm} show that GWI mentioned in Eq.(\ref{equgwi}) gives the maximum QM violation for both bipartite qutrit isotropic state and bipartite qutrit singlet state among all the GWIs derived in this paper. Henceforth, we would, therefore, consider only the GWI given by Eq.(\ref{equgwi}) in case of spin-$1$ component observables.

\section{Comparison of GWI with the CGLMP inequality for bipartite qutrits contingent upon using six-port beam splitter} 

In order to show the efficacy of GWI derived here, we will now make a comparative analysis of the QM violation obtained through GWI with that obtained by using the CGLMP inequality for bipartite qutrits using six-port beam splitters.

The CGLMP inequality \cite{cglmp} is derived based on a constraint that the correlations exhibited by a local realist theory must satisfy. This inequality has not been derived from the assumption of existence of a joint probability distribution (JPD) in the HV space. The CGLMP inequality for bipartite $3$-dimensional system has the following form
\begin{equation}
\label{cglmp}
I_3 = P(a^1=b^1)+P(b^1=a^2+1)+P(a^2=b^2)+P(b^2=a^1) - P(a^1=b^1-1)-P(b^1=a^2)-P(a^2=b^2-1)-P(b^2=a^1-1) \leq 2
\end{equation}
where, $P(a^i =  b^j + k)$ denotes the probability that the measurements $A^i$ and $B^j$ have outcomes that differ, modulo $3$, by $k$. The QM violation of the CGLMP inequality is quantified by $(I_3 - 2)$.

Here it may be noted that Wu et. al. had suggested another local realist inequality \cite{wu} which is derived based on the assumption of a local HV model satisfying the factorizability condition, using a few algebraic theorems and basic concepts of probability theory. This inequality has the following form
\begin{equation}
\label{wu}
 S = P (a^1 +, b^1 +) - P (a^1 +, b^2 +) + P (a^2 +, b^2 +) + P (a^2 0, b^1 0) + P (a^2 0, b^1 -) + P (a^2 -, b^1 0) + P (a^2 -, b^1 -) \leq 1
\end{equation}
However, an important point is that this inequality (\ref{wu}) reduces to just a version of CHSH (Clauser-Horne-Shimony-Holt) inequality \cite{CHSH} after one has grouped the outcomes ``$0$" and ``$-$" so that the inequality becomes a two outcome (``$+$" and ``not $+$") inequality. Now, it is well known that in a $2 \times 2 \times 2$ experiment (2 parties, 2 measurement settings per party, 2 outcomes per settings), all generalised Bell inequalities are simply re-writings of the CHSH inequality, obtained by linear combinations of the CHSH inequality with the appropriate normalisation conditions. Thus, the inequality (\ref{wu}) is equivalent to CHSH inequality. We will, therefore, not consider this inequality for probing efficacy of GWI for bipartite qutrits.

Before computing the effects of white noise incorporated in the states considered to GWI and CGLMP inequality, let us first obtain the maximum QM violations of CGLMP inequality for the bipartite qutrit isotropic and singlet states respectively using six-port beam splitter. If the left hand side of CGLMP inequality (\ref{cglmp}) is evaluated for isotropic state in terms of the four aforementioned trichotomic observables $a^1, a^2, b^1, b^2$ denoting observables using six-port beam splitters, then the inequality is maximally violated for the choice of measurement settings ($\phi_{a^1}(0)$, $\phi_{a^1}(1)$, $\phi_{a^1}(2)$, $\phi_{a^2}(0)$, $\phi_{a^2}(1)$, $\phi_{a^2}(2)$, $\varphi_{b^1}(0)$, $\varphi_{b^1}(1)$, $\varphi_{b^1}(2)$, $\varphi_{b^2}(0)$, $\varphi_{b^2}(1)$, $\varphi_{b^2}(1) $) = (0, 3.13, 2.64, 2.51, 4.60, 6.19, 3.73, 0.07, 1.62, 2.14, 5.81, 5.26) (in radian), and the magnitude of the maximum violation is given by  0.87293 \cite{cglmp, tt2} (Following the numerical procedure based on the downhill simplex method \cite{nm}). On the other hand, for singlet states given by Eq.(\ref{state2}), CGLMP inequality (\ref{cglmp}) is \textit{not} violated for \textit{arbitrary} choice of measurement settings.


In order to probe the efficacy of the derived GWI, we now compare the tolerances of GWI and CGLMP inequality against white noise present in a state pertaining to measurement of observables using six-port beam splitters. For this, let us consider the bipartite qutrit mixed state given by, 
\begin{eqnarray}
\label{mixedstate}
\rho = p \vert \psi \rangle \langle \psi \vert + (1-p) \frac{\mathbb{I}_3\otimes \mathbb{I}_3}{3^2} 
\end{eqnarray}
where, $p$ is the visibility parameter which changes the pure state $\vert \psi \rangle$ into a mixed state $\rho$ and $(1-p)$ denotes the amount of white noise present in the state $| \psi \rangle$ (Here we take $|\psi\rangle$ to be either the isotropic state (\ref{state1}) or the singlet state (\ref{state2})). $p=0$ denotes the maximally mixed separable state.

Now, we first consider Eq.(\ref{mixedstate}) by taking $|\psi\rangle$ as the isotropic state given by Eq.(\ref{state1}), and compute respectively the left hand side of the various local realist inequalities for the pure state $|\psi\rangle$ pertaining to measurement using six-port beam splitters. Subsequently, we repeat the computation by only taking the white noise part of Eq.(\ref{mixedstate}). After applying appropriate weightage using the visibility parameter, we obtain the various expressions of the left hand sides of the local realist inequalities corresponding to the mixed state $\rho$ mentioned in Eq.(\ref{mixedstate}) in terms of the parameter $p$. The same procedure is followed for the case of the singlet state (\ref{state2}). The minimum values of $p$ for which QM violates local realist inequalities signify the maximum amounts of white noise that can be present in the given state for the persistence of the QM violation of the relevant local realist inequality,  and this value of $p$ is known as the threshold visibility pertaining to the given local realist inequality. In the Table (\ref{tab3}), the threshold visibilities of GWI and CGLMP inequality pertaining to the bipartite qutrit isotropic and singlet states, using six-port beam splitters, are shown.

The above mentioned Table clearly shows that for the six-port beam splitter case, the CGLMP inequality given by Eq.(\ref{cglmp}) is more robust than GWI for the persistence of the QM violation in the presence of white noise incorporated in qutrit isotropic states. On the other hand, GWI given by Eq.(\ref{equgwi}) is more robust than the CGLMP inequality given by Eq.(\ref{cglmp}) for the persistence of the QM violation in the presence of white noise incorporated in qutrit singlet states. Moreover, CGLMP inequality is not violated at all by QM for qutrit singlet states using six-port beam splitter.

\begin{center}
\begin{table}
\begin{tabular}{|*{3}{c|}}
\hline
{\textit{\textbf{State}}} & \multicolumn{2}{|c|}{\textit{\textbf{Threshold visibility of}}}\\
\cline{2-3} 
 & \textit{\textbf{GWI given by Eq.(\ref{equgwi})}} & \textit{\textbf{CGLMP inequality}} \\
 \hline 
 \hline
Isotropic & $0.774$ & $0.696$ \\
\hline
Singlet & $0.774$ & $---$ \\
\hline
\end{tabular}
\caption{Threshold visibilities of GWI and CGLMP inequality for the bipartite qutrit isotropic state and singlet state using six-port beam splitter.} \label{tab3}
\end{table}
\end{center}

\section{Maximal violations of GWI and CGLMP inequality contingent upon using six-port beam splitter}

It may happen that the maximum violation of a local realist inequality is not obtained for maximally entangled states, like singlet states or isotropic states, but is obtained rather for non-maximally entangled states. One can derive the Bell operator corresponding to a local realist inequality, when observables using six-port beam splitters are measured. Any typical joint probability, say, $P(a^1=+, b^1=-)$ of obtaining outcomes $+$ and $-$ respectively, when the observable $a^1$ is measured on the first particle and the observable $b^1$ in measured on the second particle, and the initial state is $|\psi \rangle \in \mathbb{C}^{n}$, is given by,
\begin{equation}
P(a^1 =+, b^1 =-) =\langle \psi |( \{ V( \vec{\phi_{a^1}})\dagger \otimes V( \vec{\varphi_{b^1}})\dagger \} \{ |+ \rangle \langle +| \otimes |- \rangle \langle -| \} \{V( \vec{\phi_{a^1}}) \otimes V( \vec{\varphi_{b^1}})\}) | \psi \rangle
\end{equation}
where $V( \vec{\phi_{a^1}}) = U_{FT} U( \vec{\phi_{a^1}})$ and $V( \vec{\varphi_{b^1}}) = U_{FT}^* U( \vec{\varphi_{b^1}})$.
Similarly, evaluating other joint probabilities, the left hand side of any local realist inequality can be expressed for the initial state $|\psi \rangle \in \mathbb{C}^{n}$ as $\langle \psi | B | \psi \rangle$, where $B$ is the Bell operator associated with the respective local realist inequality for bipartite qutrits corresponding to using six-port beam splitter. $B$ is a $3 \otimes 3$ Hermitian Matrix. Now, for the purpose of finding the maximum eigenvalue of the Bell operator associated with a particular local realist inequality, we use the Min-Max Theorem of functional analysis and linear algebra. According to Min-Max theorem, the largest and smallest eigenvalues of a Hermitian matrix $\hat{A} \in \mathbb{C}^{n \otimes n}$ can be found as, $\lambda_{max}$ =  $ \max_{\forall x \in \mathbb{C}^{n}, x \neq 0} \frac{\langle x| \hat{A} | x \rangle}{\langle x| x \rangle}$ and $\lambda_{min}$ =  $ \min_{\forall x \in \mathbb{C}^{n}, x \neq 0} \frac{\langle x| \hat{A} | x \rangle}{\langle x| x \rangle}$ respectively.

Using the above mentioned procedure it is found that, contingent upon using six-port beam splitter, the maximum QM violation of GWI given by Eq.(\ref{equgwi}) is 0.20711, which is larger than the maximum QM violations of GWI given by Eq.(\ref{equgwi}) for bipartite qutrit isotropic and singlet states. Its corresponding eigenvector is a non-maximally entangled state of two qutrits, which has the following form
\begin{equation} 
\label{mvgw}
| \psi_{gwi} \rangle = -0.35| 00 \rangle + 0.35| 01 \rangle + 0.09| 02 \rangle + 0.35| 10 \rangle  -0.35 | 11 \rangle - 0.09 | 12 \rangle + 0.09 | 20 \rangle - 0.09 | 21 \rangle + 0.70 | 22 \rangle
\end{equation}
Therefore, the threshold visibility of GWI given by Eq.(\ref{equgwi}) for the state given by Eq.(\ref{mvgw}) is 0.682.

The maximum QM violation of CGLMP inequality given by Eq.(\ref{cglmp}) is 0.9149 \cite{tt2}, which is a bit larger than the maximum QM violations of CGLMP inequality for bipartite qutrit isotropic and much larger than that of qutrit singlet states. Its corresponding eigenvector is a non-maximally entangled state of two qutrits, which has the following form
\begin{equation} 
\label{mvcg}
| \psi_{c} \rangle = \frac{1}{\sqrt{2 + (0.792)^2}} ( | 00 \rangle + (0.792) |11 \rangle + |22 \rangle)
\end{equation}
The threshold visibility of CGLMP inequality given by Eq.(\ref{cglmp}) for the state given by Eq.(\ref{mvcg}) is 0.686.

Hence the maximum threshold visibility, contingent upon using six-port beam splitter, of GWI given by Eq.(\ref{equgwi}) corresponding to a non-maximally entangled state is smaller than that of CGLMP inequality.

\section{Comparison of GWI with the CGLMP inequality for bipartite qutrits contingent upon using spin-$1$ component observables}

Now, again in order to show the efficacy of the derived GWI, contingent upon using spin-$1$ component observables, we will now make a comparative analysis of the QM violation obtained through GWI with that obtained by CGLMP inequality. We perform the required comparison by inserting white noise to the pure isotropic (mentioned in Eq.(\ref{state1})) and singlet (mentioned in Eq.(\ref{state2})) states.


Before computing the effect of white noise incorporated in the states considered, let us first obtain the maximum QM violations of CGLMP inequality for the bipartite qutrit isotropic and singlet states  respectively contingent upon using spin-$1$ component observables. If LHS of the CGLMP inequality (\ref{cglmp}) is evaluated in terms of the four trichotomic observables $a^1, a^2, b^1, b^2$  denoting spin-$1$ components in arbitrary directions for the isotropic state, then (from numerical procedure based on the downhill simplex method \cite{nm}) it can be shown that for the choice of measurement settings ($\theta_1^a,\phi_1^a;\theta_2^a,\phi_2^a;\theta_1^b,\phi_1^b;\theta_2^b,\phi_2^b$) = (0.45, 6.28; 1.35, 6.28; 0, 1.07; 0.90, 6.28) (in radian), CGLMP inequality is maximally violated and the magnitude of the maximum violation is given by 0.52951. On the other hand, it can be shown that if the measurements of trichotomic observables $a^1, a^2, b^1, b^2$ denoting spin-$1$ components in arbitrary directions are performed on $3\otimes 3$-dimensional singlet state given by, Eq.(\ref{state2}), then (from numerical procedure based on the downhill simplex method \cite{nm}) it can be shown that the maximum QM violation of the CGLMP inequality is  0.52951. This occurs for the choice of measurement settings ($\theta_1^a,\phi_1^a;\theta_2^a,\phi_2^a;\theta_1^b,\phi_1^b;\theta_2^b,\phi_2^b$) = (0.78, 5.72; 0.88, 4.47; 2.12, 3.08; 2.41, 1.92)  (in radian). 

In order to probe the efficacy of the derived GWI, we now compare the tolerances of GWI against white noise present in a state with that of CGLMP inequality, contingent upon using spin-$1$ component observables, in a similar way described in Section VI.  In the Table (\ref{tab5}), the threshold visibilities of GWI and CGLMP inequality pertaining to the bipartite qutrit isotropic and singlet states respectively are shown.

The above mentioned Table clearly shows that the GWI given by Eq.(\ref{equgwi}) is more robust than the CGLMP inequality for the persistence of the QM violation in the presence of white noise incorporated in both the qutrit isotropic and singlet states.

\begin{center}
\begin{table}
\begin{tabular}{|*{3}{c|}}
\hline
{\textit{\textbf{State}}} & \multicolumn{2}{|c|}{\textit{\textbf{Threshold visibility of}}}\\
\cline{2-3} 
 & \textit{\textbf{GWI given by Eq.(\ref{equgwi})}} & \textit{\textbf{CGLMP inequality}} \\
 \hline 
 \hline
Isotropic & $0.786$ & $0.791$ \\
Singlet & $0.786$ & $0.791$ \\
\hline
\end{tabular}
\caption{Threshold visibilities of GWI and CGLMP inequality for the bipartite qutrit isotropic state and singlet state using spin-$1$ component observables.} \label{tab5}
\end{table}
\end{center}

\section{Maximal violations of GWI and CGLMP inequality contingent upon using spin-$1$ component observables}

As discussed in Section VII, the maximum QM violations of GWI and CGLMP inequality using spin-$1$ component observables can be evaluated using the Min-Max theorem as stated before. Here any typical joint probability, say, $P(a^1=+, b^1=-)$ of obtaining outcomes $+$ and $-$ respectively, when the observable $a^1$ is measured on the first particle and the observable $b^1$ in measured on the second particle, and the initial state is $|\psi' \rangle \in \mathbb{C}^{n}$, is given by,
\begin{equation}
P(a^1=+, b^1=-) =\langle \psi' |( \vert + \rangle_{(\theta_{1},\phi_{1})} \langle + \vert \otimes \vert - \rangle_{(\theta_{3},\phi_{3})} \langle - \vert) | \psi' \rangle
\end{equation}
Similarly, evaluating other joint probabilities, the LHS of any local realist inequality can be expressed for the initial state $|\psi' \rangle \in \mathbb{C}^{n}$ as $\langle \psi' | B' | \psi' \rangle$, where $B'$ is the Bell operator associated with the respective local realist inequality for bipartite qutrits corresponding to using spin-$1$ component observables. $B'$ is a $3 \otimes 3$ Hermitian Matrix. The largest eigenvalue of $B'$ will be the maximum QM violation of the corresponding local realist inequality using spin-$1$ component observables and using the aforementioned Min-Max theorem, one can find the largest eigenvalues of the Bell operators associated with different local realist inequalities for bipartite qutrits.

We have found that the maximum QM violation of GWI given by Eq.(\ref{equgwi}) using spin-$1$ component observables is 0.20711, which is larger than the maximum QM violations of GWI given by Eq.(\ref{equgwi}) for bipartite qutrit isotropic and singlet states. Its corresponding eigenvector is a non-maximally entangled state of two qutrits, which has the following form
\begin{equation} 
\label{statemvg}
| \psi'_{gwi} \rangle = -0.01| 00 \rangle - 0.01 | 01 \rangle + 0.67 | 02 \rangle - 0.18 | 10 \rangle - 0.40 | 11 \rangle - 0.19 | 12 \rangle + 0.23 | 20 \rangle + 0.51 | 21 \rangle - 0.13 | 22 \rangle
\end{equation}
Therefore, the threshold visibility of GWI given by Eq.(\ref{equgwi}) for the state given by Eq.(\ref{statemvg}) is 0.682.

The maximum QM violation of CGLMP inequality given by Eq.(\ref{cglmp}) using spin-$1$ component observables is 0.62877, which is a bit larger than the maximum QM violations of CGLMP inequality for bipartite qutrit isotropic and singlet states. Its corresponding eigenvector is a non-maximally entangled state of two qutrits, which has the following form
\begin{equation} 
| \psi'_{c} \rangle = (0.51 - 0.15 i)| 00 \rangle - (0.22 + 0.28 i)| 01 \rangle  + (0.05 + 0.13 i)| 02 \rangle  -(0.28 + 0.12 i)| 10 \rangle - (0.11 + 0.31 i)| 11 \rangle  \nonumber
\end{equation}
\begin{equation}
\label{statemvc}
- (0.12 - 0.23 i)| 12 \rangle  +  0.21 i | 20 \rangle - (0.17 - 0.22 i)| 21 \rangle - 0.42 | 22 \rangle
\end{equation}
The threshold visibility of CGLMP inequality given by Eq.(\ref{cglmp}) for the state given by Eq.(\ref{statemvc}) is 0.761.

Hence the maximum threshold visibility, contingent upon using spin-$1$ component observables, of GWI given by Eq.(\ref{equgwi}) corresponding to a non-maximally entangled state is much smaller than that of CGLMP inequality.

To summarize, while the maximum QM violation of GWI occurs for the state given by Eq.(\ref{statemvg}), maximum QM violation of CGLMP inequality occurs for the state given by Eq.(\ref{statemvc}) when one uses spin-$1$ component observables. Note that none of the states is a maximally entangled state. Interestingly, the maximum QM violations of GWI given by Eq.(\ref{equgwi}) are the same whether one uses spin-$1$ component observables or six-port beam splitter. But the maximum QM violations of CGLMP inequality given by Eq.(\ref{cglmp}) differ for the two different types of observables stated above.

\section{QM violation of GWI by mixed bipartite qutrit states}

\begin{figure}[t]
\centering
\begin{minipage}{.45\textwidth}
  \centering
  \begin{framed}
  \includegraphics[width=7.5cm,height=6cm]{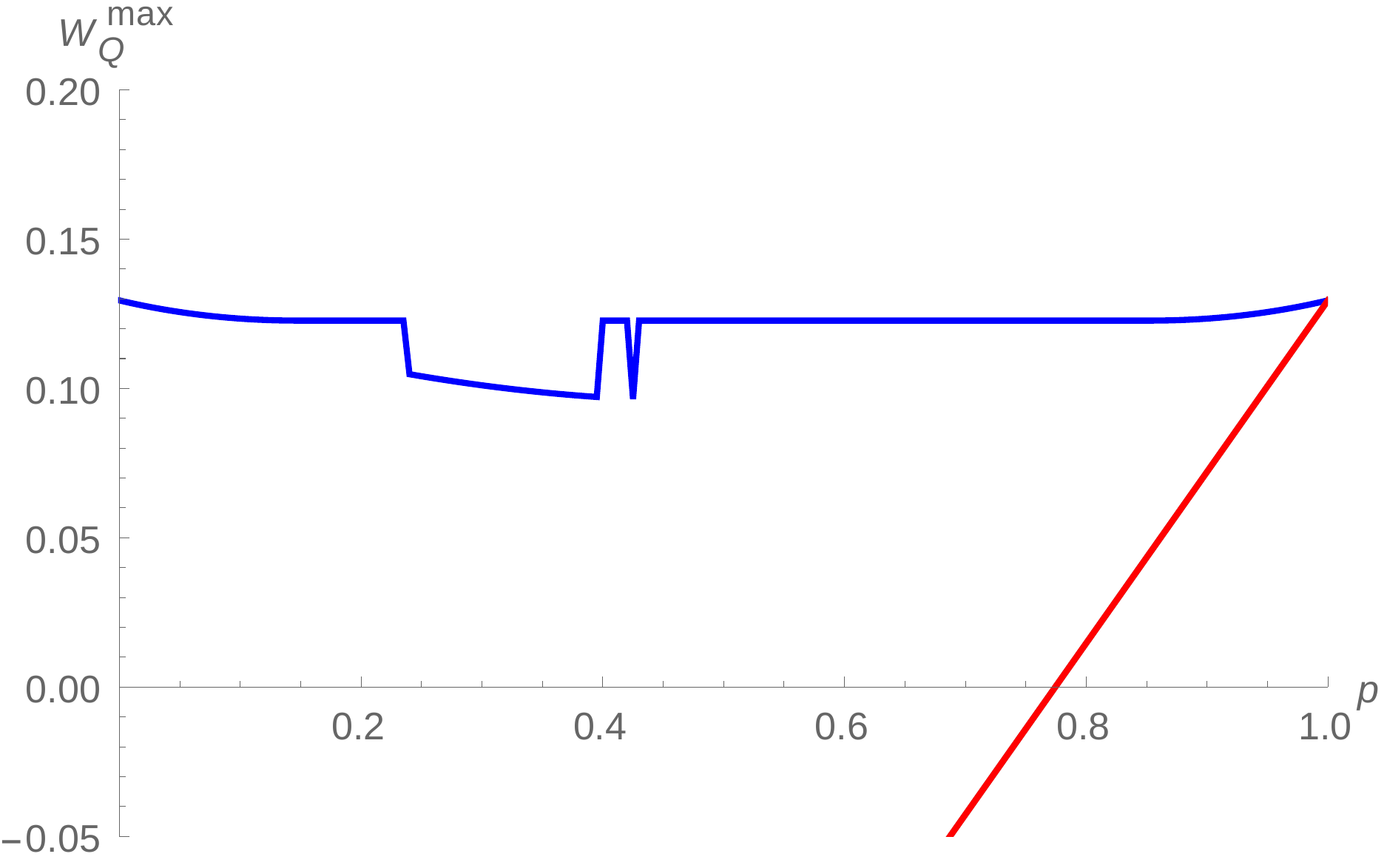}
  \captionof{figure}{Red plot represents the $W_{Q}^{max}$ versus $p$ for $\rho_1$ and $\rho_2$ using six-port beam splitter. Blue plot represents the $W_{Q}^{max}$ versus $p$ for $\rho_3$ using six-port beam splitter.}
  \label{fig1}
  \end{framed}
\end{minipage}%
\begin{minipage}{.45\textwidth}
\begin{framed}
  \centering
  \includegraphics[width=7.5cm,height=5.6cm]{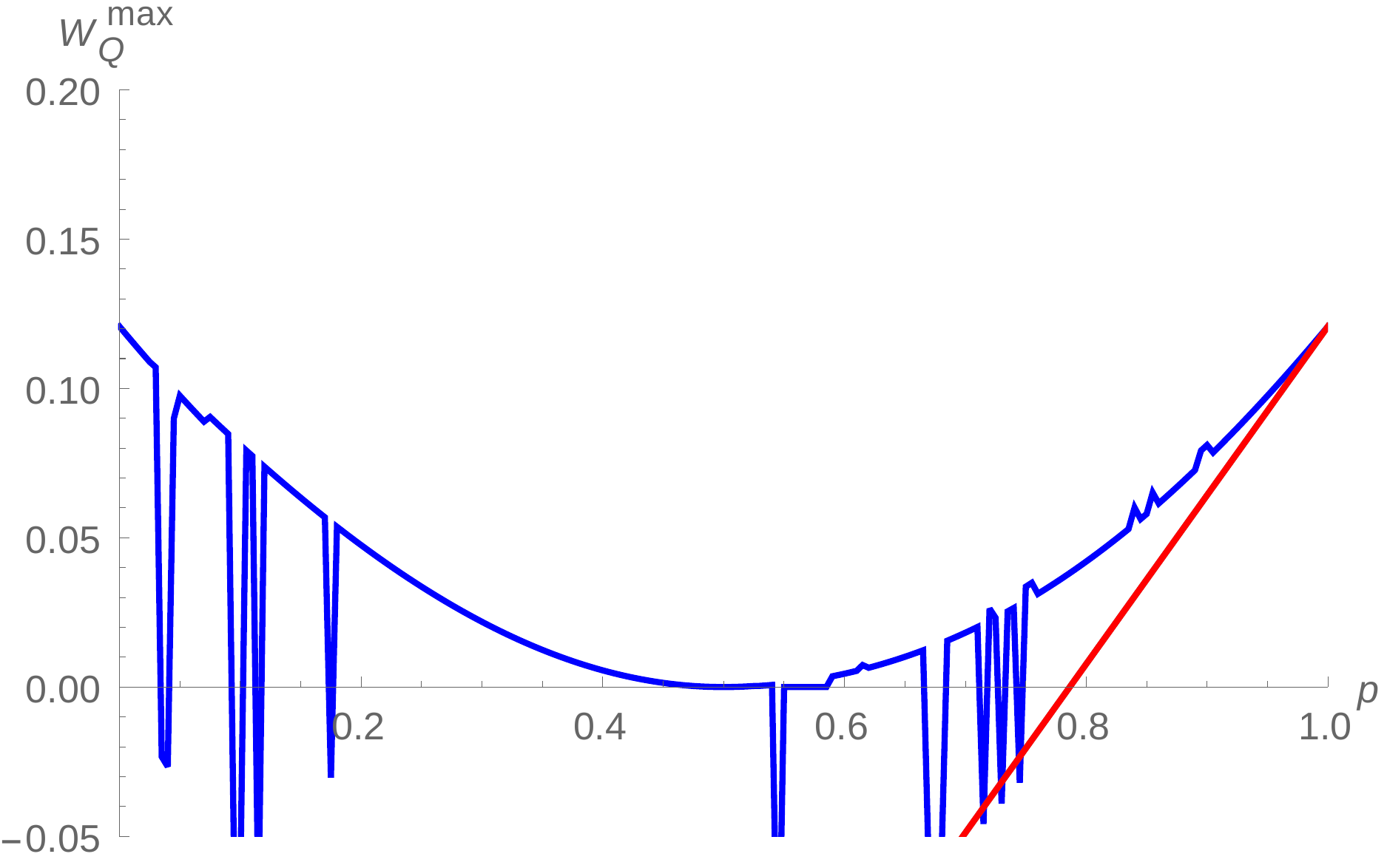}
  \captionof{figure}{Red plot represents the $W_{Q}^{max}$ versus $p$ for $\rho_1$ and $\rho_2$ using spin-$1$ component observables. Blue plot represents the $W_{Q}^{max}$ versus $p$ for $\rho_3$ using spin-$1$ component observables.}
  \label{fig2}
  \end{framed}
\end{minipage}
\end{figure}

It is well known that all pure bipartite entangled states violate local realist inequalities \cite{gisin}. This is, however, not true for mixed states. Hence, the relation between entanglement and QM violations of local realist inequalities for mixed states is not clear. The connection between entanglement and mixedness of the state, and the amount of QM violation of different local realist inequalities by that state is another area of interest. There are instances demonstrating that to produce an equal amount of Bell-CHSH (Bell-Clauser-Horne-Shimony-Holt) violation some states require more entanglement than others. In Ref. \cite{mixed1}, it was suggested that with the increase in mixedness of bipratite qubit state, higher degree of entanglement is required for it to violate the Bell-CHSH inequality. However, there are counter-examples showing the existence of states with equal amount of Bell-CHSH violation and entanglement, but one of them is more mixed than the other. The reason as to why equal amount of Bell-CHSH violation requires different amounts of entanglement cannot be explained by mixedness alone \cite{mixed2}. For a class of bipartite qubit mixed states it has been shown that it is not possible to discriminate between states violating or not violating Bell-CHSH inequalities, knowing only their entanglement and mixedness \cite{mixed3}.

In this Section we have investigated the QM violations of GWI by four different classes of mixed bipartite qutrit states which are given by
\begin{equation}
\rho_1 = p \vert \psi_1 \rangle \langle \psi_1 \vert + (1-p) \frac{\mathbb{I}_3\otimes \mathbb{I}_3}{3^2} 
\end{equation}
where $\vert\psi_1\rangle = \frac{\vert 00 \rangle + \vert 11 \rangle + \vert 22 \rangle}{\sqrt{3}}$ is the bipartite qutrit pure isotropic states; $0 \leq p \leq 1$.
\begin{equation}
\rho_2 = p \vert \psi_2 \rangle \langle \psi_2 \vert + (1-p) \frac{\mathbb{I}_3\otimes \mathbb{I}_3}{3^2} 
\end{equation}
where $\vert\psi_2\rangle = \frac{\vert 02 \rangle - \vert 11 \rangle + \vert 20 \rangle}{\sqrt{3}}$ is the bipartite qutrit pure singlet states; $0 \leq p \leq 1$.
\begin{equation}
\rho_3 = p \vert \psi_1 \rangle \langle \psi_1 \vert + (1-p) \vert \psi_2 \rangle \langle \psi_2 \vert
\end{equation}
where $\vert\psi_1\rangle$ and $\vert\psi_2\rangle$ are the bipartite qutrit pure isotropic state and singlet state respectively; $0 \leq p \leq 1$.
\begin{equation}
\rho_4 = p\Big( q \vert \psi_1 \rangle \langle \psi_1 \vert + (1-q) \vert \psi_2 \rangle \langle \psi_2 \vert \Big) + (1-p) \frac{\mathbb{I}_3\otimes \mathbb{I}_3}{3^2} 
\end{equation}
where $\vert\psi_1\rangle$ and $\vert\psi_2\rangle$ are the bipartite qutrit pure isotropic state and singlet state respectively; $0 \leq p \leq 1$; $0 \leq q \leq 1$.

\begin{figure}[t]
\centering
\begin{minipage}{.45\textwidth}
  \centering
  \begin{framed}
  \includegraphics[width=7.5cm,height=6cm]{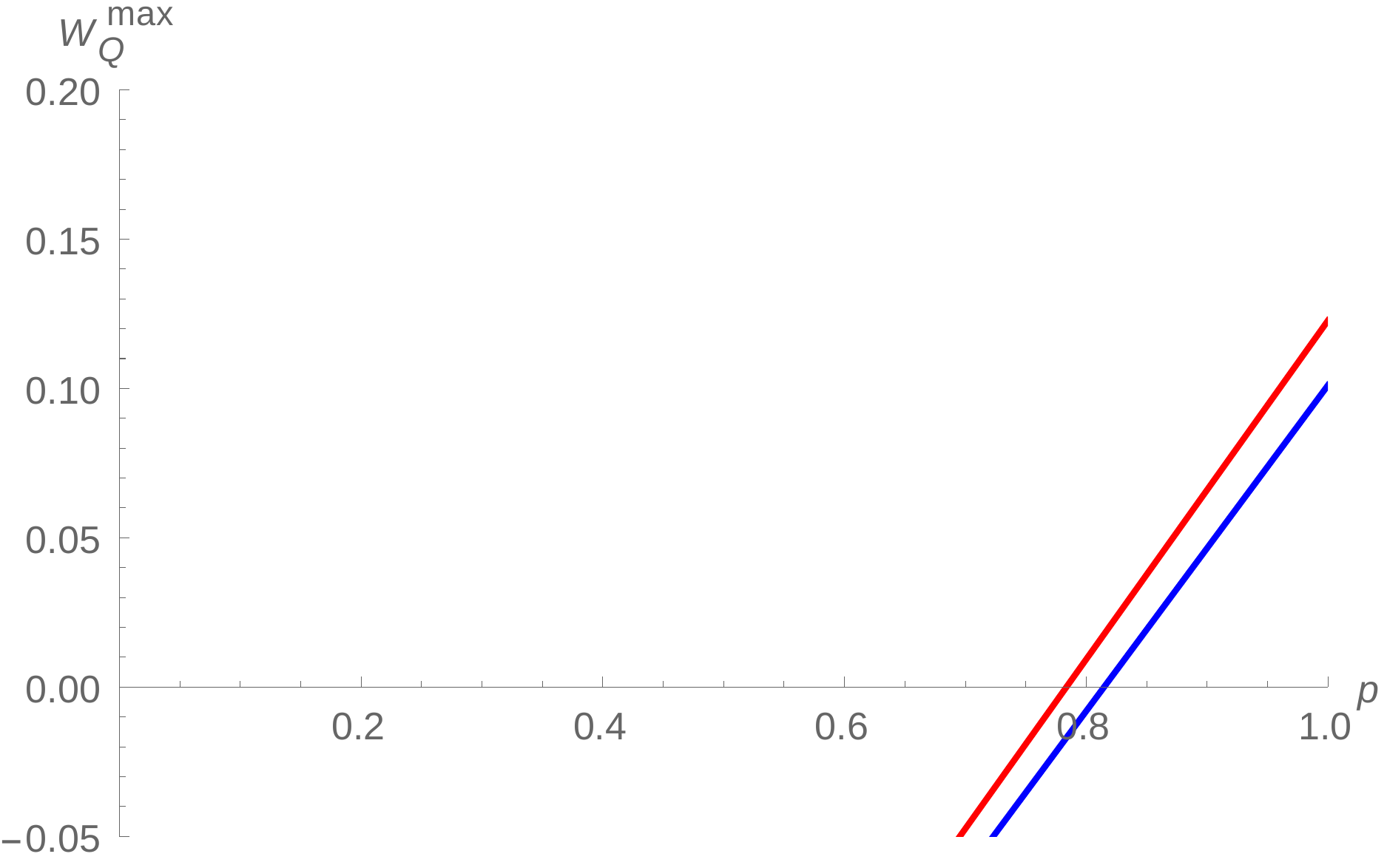}
  \captionof{figure}{Red and blue plot represent the $W_{Q}^{max}$ versus $p$ for $\rho_4$ with $q=0.7$ and with $q=0.3$, respectively, using six-port beam splitter.}
  \label{fig3}
  \end{framed}
\end{minipage}%
\begin{minipage}{.45\textwidth}
\begin{framed}
  \centering
  \includegraphics[width=7.5cm,height=5.6cm]{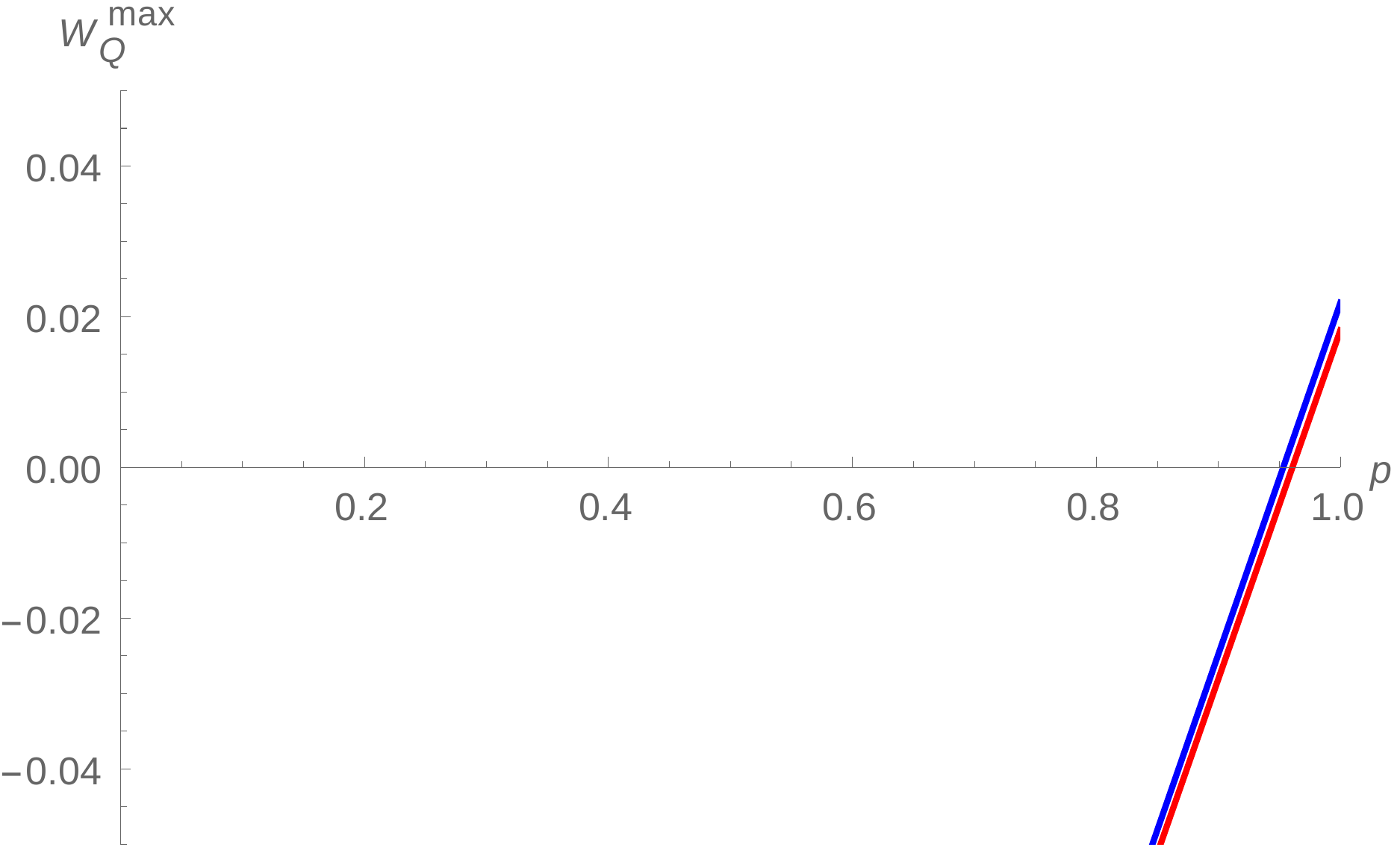}
  \captionof{figure}{Red and blue plot represent the $W_{Q}^{max}$ versus $p$ for $\rho_4$ with $q=0.7$ and with $q=0.3$, respectively, using spin-$1$ component observables..}
  \label{fig4}
  \end{framed}
\end{minipage}
\end{figure}

Varying over measurement settings, we have calculated numerically (based on the downhill simplex method \cite{nm}) the maximum QM violations of GWI given by Eq.(\ref{equgwi}) using six-port beam splitter and by using spin-$1$ component observables for different values of the state parameters. In Fig.(\ref{fig1}) and Fig.(\ref{fig2}) we have plotted the maximum values of the left hand side of GWI given by Eq.(\ref{equgwi}) ($W^{max}_{Q}$) for the states $\rho_1$, $\rho_2$ and $\rho_3$ for different values of the state parameter $p$ using six-port beam splitter and using spin-$1$ component observables respectively. In Fig.(\ref{fig3}) and Fig.(\ref{fig4}) we have plotted the maximum values of the left hand side of GWI given by Eq.(\ref{equgwi}) ($W^{max}_{Q}$) for the states $\rho_4$ for $q=0.3$ and $q=0.7$ for different values of the state parameter $p$ using six-port beam splitter and using spin-$1$ component observables respectively. For other values of $q$ the plot is similar, but we have not shown them in the Figures.

Fig.(\ref{fig1}), Fig.(\ref{fig2}), Fig.(\ref{fig3}) and Fig.(\ref{fig4}) show that each of         $\rho_1$, $\rho_2$ and $\rho_4$ (with $q=0.3$ and $q=0.7$) violates the GWI given by Eq.(\ref{equgwi}) above a particular value of $p$ using six-port beam splitter as well as using spin-$1$ component observable. Above this particular value of $p$, the magnitude of QM violation of GWI by each of these states increases linearly with increasing values of $p$. Moreover, the magnitudes of QM violations of GWI for $\rho_1$ and $\rho_2$ are the same for any value of $p$.

Fig.(\ref{fig1}) indicates that the state $\rho_3$ violates the GWI given by Eq.(\ref{equgwi}) using six-port beam splitter for any values of $p$. However, the magnitude of the QM violation of GWI by the state $\rho_3$ is not a linear function of $p$.  Fig.(\ref{fig2}) indicates that the state $\rho_3$ violates the GWI given by Eq.(\ref{equgwi}) using spin-$1$ component observable in some particular range of $p$. In this case, the magnitude of the QM violation of GWI by the state $\rho_3$ is not a linear function of $p$.

\section{Conclusion}

In this work we have extended Wigner's approach \cite{Wigner} by deriving generalized Wigner type local realist inequalities (GWI) for bipartite qutrit systems based on the assumption of the existence of the overall joint probability distributions in the underlying stochastic HV space for the measurement outcomes pertaining to the relevant trichotomic observables, that satisfy the locality condition, and yield the measurable marginal probabilities. An important point to stress here is that the expressions of GWIs that have been derived here do \textit{not} reduce to that of Bell-CHSH inequalities by grouping any two outcomes. This feature distinguishes GWIs considered here from the other local realist inequalities for bipartite qutrits; for example, the one suggested by Wu et. al. \cite{wu}. Also, note that the factorizability condition for hidden variables used in deriving Bell-CHSH inequalities is not required for deriving GWI. In this context, it should be mentioned that the role of factorizability condition for stochastic hidden variables has been subjected to a critical examination \cite{Alex}. Our work based on GWI serves to validate the notion that assuming the existence of overall joint probabilities in any stochastic HV theory yielding the measurable marginal probabilities is sufficient to demonstrate for the bipartite qutrit systems an incompatibility between QM and a class of stochastic HV theories satisfying the locality condition.

Efficacy of the derived GWI for the bipartite qutrit systems has been probed by analysing the robustness of its QM violation against white noise incorporated in the states considered. A comparative study of GWI in these contexts has been performed with respect to CGLMP inequality \cite{cglmp}, using the two widely used maximally entangled states, {\it viz.} the singlet state, and the isotropic state which are also considered to be relevant in quantum information processing. Using six-port beam splitter, we have found that CGLMP inequality has lower threshold visibility compared to GWI when white noise is incorporated in the qutrit isotropic states. On the other hand, GWI has lower threshold visibility compared to CGLMP inequality when white noise is introduced in qutrit singlet states. In fact, CGLMP inequality is not violated by QM for qutrit singlet states when six-port beam splitter is used. It is also found that, contingent upon using spin-$1$ component observables, GWI has a lower threshold visibility than CGLMP inequality when white noise is introduced in both isotropic and singlet states. We can, therefore, state that for showing QM incompatibility of qutrit \textit{isotropic} states with local realism using six-port beam splitter, the CGLMP inequality is more efficient in terms of the robustness of its QM violation against white noise than GWI; whereas, for qutrit \textit{singlet} states, GWI is more efficient than the CGLMP inequality in showing QM incompatibility with local realism using six-port beam splitter. On the other hand, if one uses spin-$1$ component observables, GWI is more efficient in showing QM incompatibility with local realism for \textit{both} qutrit isotropic and singlet states in the presence of white noise in these states, compared to the CGLMP inequality.

Another significant result obtained in this paper is that, for both six-port beam splitter and spin-$1$ component observables, the maximum QM violations of GWI and the CGLMP inequality occur for non-maximally entangled states. Further, it is found that the maximum QM violation of GWI is the same whether one uses spin-$1$ component observables or the observables pertaining to the six-port beam splitter. On the other hand, the maximum QM violation of the CGLMP inequality differs for the two different types of observables stated above.

QM violations of GWI for different classes of mixed bipartite qutrit states using the two aforementioned types of observables have also been addressed.

It requires to be studied what interesting results such comparison between GWI and the CGLMP inequality would yield when extended in the context of non maximally entangled bipartite qutrit pure and mixed states. Finally, it should be worth probing the possibility of any information theoretic application of GWI similar to that of the more familiar Bell-CHSH inequalities, for example, in the context of device independent quantum key generation \cite{diqkd}, and for developing robust multipartite multilevel quantum protocols \cite{ns}.

\begin{acknowledgements}
DD acknowledges the financial support from University Grants Commission (UGC), Govt. of India.   DH, ASM and SG acknowledge support from the project SR/S2/LOP-08/2013 of DST, India. The research of DH is also supported by the Center for Science, Kolkata. SD acknowledges support from DST-INSPIRE fellowship.
\end{acknowledgements}

\appendix
\section{Impossibility to reduce the GWI for bipartite qutrits to equivalent classes of Bell-CHSH inequalities}	
In this section, we are going to justify that the GWI for bipartite qutrits derived in this paper cannot be reduced to equivalent classes of Bell-CHSH inequalities.

The left hand sides of the GWI given by (\ref{equgwi2}) and (\ref{equgwi3}) are linear combinations of joint probabilities of the form $p(a^k = m_i, b^l = m_j)$ ($k,l=1,2;$ $i,j=1,2,3$). \textit{If} the GWI given by (\ref{equgwi2}) and (\ref{equgwi3}) can be reduced to local realist inequalities in the $2\times 2 \times 2$ experimental scenario ($2$ parties, $2$ measurement settings per party, $2$ outcomes per measurement setting) by grouping the outcomes ``$m_2$" and ``$m_3$", then the left hand side of each GWI would have to be necessarily a linear combination of the probability distributions $p(a^i = m_1, b^j = m_1)$, $p(a^k = m_1, b^l = \overline{m_1})$, $p(a^q = \overline{m_1}, b^r = m_1)$ and $p(a^s = \overline{m_1}, b^t = \overline{m_1})$ ($i,j,k,l,q, r, s, t = 1, 2$; ``$\overline{m_1}$" implies ``Not $m_1$") in the following form,
\begin{equation}
\sum_{i=1}^{2} \sum_{j=1}^{2} \Big[ q_{ij} p(a^i = m_1, b^j = m_1) +  r_{ij} p(a^i = m_1, b^j = \overline{m_1})+  s_{ij} p(a^i = \overline{m_1}, b^j = m_1) + t_{ij} p(a^i = \overline{m_1}, b^j = \overline{m_1}) \Big] 
\end{equation}
where $q_{ij}$, $r_{ij}$, $s_{ij}$, $t_{ij}$ ($i,j=1,2$) are constants and
\begin{equation}
p(a^k = m_1, b^l = \overline{m_1}) = \sum_{x=2,3} p(a^k = m_1, b^l = m_x) 
\end{equation}
\begin{equation}
p(a^q = \overline{m_1}, b^r = m_1) = \sum_{x=2,3} p(a^q = m_x, b^r = m_1) 
\end{equation}
\begin{equation}
p(a^s = \overline{m_1}, b^t = \overline{m_1}) = \sum_{x=2,3} \sum_{y=2,3} p(a^s = m_x, b^t = m_y)
\end{equation}

Hence, to make the grouping of the outcomes ``$m_2$" and ``$m_3$" possible, the left hand side of each GWI should have the form
\begin{equation}
\sum_{i=1}^{2} \sum_{j=1}^{2} \Big[ q_{ij} p(a^i = m_1, b^j = m_1)  +  r_{ij} \sum_{x=2,3} p(a^i = m_1, b^j = m_x) +  s_{ij}  \sum_{x=2,3} p(a^i = m_x, b^j = m_1) \nonumber
\end{equation} 
\begin{equation}
\label{lc}
 +  t_{ij}  \sum_{x=2,3} \sum_{y=2,3} p(a^i = m_x, b^j = m_y) \Big]
\end{equation} 
Now, note the following salient points:\\
$\bullet$ In the left hand side of the GWI given by (\ref{equgwi2}) the coefficients of $p(a^1=m_1, b^1 = m_2)$ and $p(a^1=m_1, b^1 = m_3)$ are $1$ and $0$ respectively. On the other hand, both the coefficients of $p(a^1=m_1, b^1 = m_2)$ and $p(a^1=m_1, b^1 = m_3)$ in Eq.(\ref{lc}) are equal to  $r_{11}$.\\
$\bullet$  Again, the coefficients of $p(a^2=m_1, b^1 = m_2)$ and $p(a^2=m_1, b^1 = m_3)$ in the left hand side of the GWI (\ref{equgwi2}) are $-1$ and $0$ respectively. On the other hand, both the coefficients of $p(a^2=m_1, b^1 = m_2)$ and $p(a^2=m_1, b^1 = m_3)$ in Eq.(\ref{lc}) are  equal to   $r_{21}$.\\ 
$\bullet$ The coefficients of $p(a^1=m_1, b^2 = m_2)$ and $p(a^1=m_1, b^2 = m_3)$ in the left hand side of the GWI (\ref{equgwi2}) are $-1$ and $0$ respectively. On the other hand, both the coefficients of $p(a^1=m_1, b^2 = m_2)$ and $p(a^1=m_1, b^2 = m_3)$ in Eq.(\ref{lc}) are  equal to   $r_{12}$.\\
$\bullet$ The coefficients of $p(a^2=m_2, b^2 = m_2)$, $p(a^2=m_2, b^2 = m_3)$, $p(a^2=m_3, b^2 = m_2)$ and $p(a^2=m_3, b^2 = m_3)$ in the left hand side of the GWI (\ref{equgwi2}) are $0$, $-1$, $0$ and $-1$ respectively. On the other hand, the coefficients of $p(a^2=m_2, b^2 = m_2)$, $p(a^2=m_2, b^2 = m_3)$, $p(a^2=m_3, b^2 = m_2)$ and $p(a^2=m_3, b^2 = m_3)$ in Eq.(\ref{lc}) are equal  to $t_{22}$.

Hence, the left hand side of the GWI given by (\ref{equgwi2}) does not have the form of Eq.(\ref{lc}). In a similar way, it can be shown that the left hand side of the GWI given by (\ref{equgwi3}) does not have the form given by Eq.(\ref{lc}).

It is, therefore, impossible to reduce the GWI given by (\ref{equgwi2}) and (\ref{equgwi3}) to local realist inequalities in $2\times 2 \times 2$ experimental scenario by grouping the outcomes ``$m_2$" and ``$m_3$". In a similar way, it can be shown that it is impossible to reduce the GWI given by (\ref{equgwi2}) and (\ref{equgwi3}) to local realist inequalities in $2\times 2 \times 2$ experimental scenario by grouping the outcomes ``$m_1$" and ``$m_2$" or, by grouping the outcomes ``$m_1$" and ``$m_3$".


Now, all generalised Bell inequalities in $2 \times 2 \times 2$ experimental scenario are simply re-writings of the Bell-CHSH (Bell-Clauser-Horne-Shimony-Holt) inequalities, obtained by linear combinations of the Bell-CHSH inequalities with the normalisation condition and the no-signalling condition. Since, the GWI given by (\ref{equgwi2}) and (\ref{equgwi3}) cannot be reduced to local realist inequalities in $2\times 2 \times 2$ experimental scenario by grouping any two outcomes, it follows that the GWI (\ref{equgwi2}) and (\ref{equgwi3}) cannot be reduced to equivalent classes of Bell-CHSH inequalities.





\end{document}